\documentclass[prb,final,twocolumn,superscriptaddress,showpacs]{revtex4}
\usepackage{graphicx}

\begin{document}

\title{Dual electronic states in thermoelectric cobalt oxide}

\author{Patrice Limelette}
\affiliation{Laboratoire LEMA, UMR 6157 CNRS-CEA, Universit\'e F. Rabelais, UFR Sciences, Parc de Grandmont, 37200 Tours, France}
\author{Sylvie H\'ebert}
\affiliation{Laboratoire CRISMAT, UMR CNRS--ENSICAEN(ISMRA) 6508, 6, Bld du Mar\'echal Juin, 14050 CAEN Cedex, France}
\author{Herv\'e Muguerra}
\affiliation{Laboratoire CRISMAT, UMR CNRS--ENSICAEN(ISMRA) 6508, 6, Bld du Mar\'echal Juin, 14050 CAEN Cedex, France}
\author{Raymond Fr\'esard}
\affiliation{Laboratoire CRISMAT, UMR CNRS--ENSICAEN(ISMRA) 6508, 6, Bld du Mar\'echal Juin, 14050 CAEN Cedex, France}
\author{Charles Simon}
\affiliation{Laboratoire CRISMAT, UMR CNRS--ENSICAEN(ISMRA) 6508, 6, Bld du Mar\'echal Juin, 14050 CAEN Cedex, France}

\date{\today}

\begin{abstract}
\vspace{0.3cm}
We investigate the low temperature magnetic field dependence of the resistivity in the thermoelectric misfit cobalt oxide 
[Bi$_{1.7}$Ca$_{2}$O$_{4}$]$_{0.59}$CoO$_{2}$ from 60\, K down to 3\, K. 
The scaling of the negative magnetoresistance demonstrates a spin dependent transport mechanism due to a strong Hund's coupling. 
The inferred microscopic description implies dual electronic states which explain the coexistence between localized and itinerant electrons 
both contributing to the thermopower. By shedding a new light on the electronic states which lead to a high thermopower, this result likely provides a new potential way to optimize the thermoelectric properties.
\end{abstract}

\pacs{72.15.Jf 71.27.+a 72.25.-b}

\maketitle

\section{introduction}

By converting heat into voltage, thermoelectric materials are not only of major interest in both energy saving and 
cooling applications but they also bring a fundamental challenge in order to find the physical limits optimizing their performance.
As stated by Mahan and Sofo, \cite{mahan96} this problem can be formulated as follows: 
\textit{What is the best electronic structure a thermoelectric can have?}
In their answer they stress that the energy distribution of the charge carriers should be as narrow as possible, \cite{mahan96}
thus  emphasizing the relevance of correlated materials to thermoelectricity as those in the vicinity of a Mott 
metal-insulator transition. \cite{Palsson98,Merino00,Georges96}
In particular, layered cobaltates  seem to belong to this class of materials:\cite{LimelettePRB05} they exhibit interesting properties, \cite{Terasaki97} including superconductivity\cite{Takada03} and both metalliclike resistivity 
and large thermopower at room temperature.
In particular, it has been recently pointed out that their large thermopower
could both result from extended quasiparticles with an enhanced 
effective mass, and from an entropy contribution of localized
spins. \cite{LimelettePRL06} Indeed the room temperature thermopower is large
in the whole series of cobalt
misfits and does not depend  on the nature of the separating
layers. \cite{Bobroff07} In misfits, it ranges from 75\, $\mu$V/K
in [Sr$_2$O$_{2-\delta}$]$_{0.53}$ [CoO$_2$] as observed by Ishiwata et al.,
\cite{Ishiwata06} to 350\, $\mu$V/K in (CaOH)$_{1.14}$CoO$_2$, 
\cite{Shizuya07} the former value being close to the expected value for an
assembly of free spins 1/2. On the other hand, a lack of systematic trend has
been shown for both resistivity and thermopower of these systems at low
temperature. \cite{Bobroff07} The low temperature thermopower exhibits a
strong magnetic field dependence in BiCaCoO and in Na$_x$CoO$_2$,
\cite{Wang03} but not in BiBaCoO. \cite{Hervieu2003} 
While for most cobaltates, there is a large temperature range where the thermopower is T-linear, this behavior does not extrapolate to zero at zero temperature 
in BiCaCoO, in contrast to, \textit{e.g.}, BiBaCoO. \cite{Maignan03} 
Furthermore, it is known that the entropic contribution\cite{Koshibae00,Koumoto02,Maignan03,Wang03} 
or the electronic correlations\cite{Palsson98,Merino00,behnia04,Motrunich03,LimelettePRB05,Fresard02} can increase thermopower, 
these two additive contributions further raise the aforementioned issue:
what is the nature of the electronic states relevant to the transport?

\begin{figure}[htbp]
\centerline{\includegraphics[width=0.6\hsize]{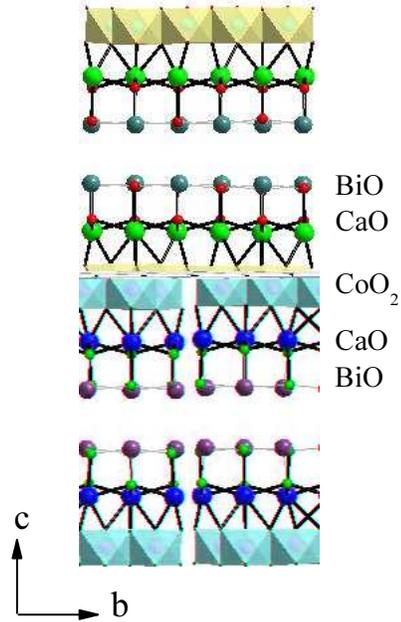}}
\caption{(color online). Schematic picture of the crystal structure of the cobaltate [Bi$_{1.7}$Ca$_{2}$O$_{4}$]$_{0.59}$CoO$_{2}$.}
\label{figbicacoo}
\end{figure}

To that aim we report in this paper on a crossed experimental
investigation of the magnetoresistance and thermopower in single crystal cobalt oxide 
[Bi$_{1.7}$Ca$_{2}$O$_{4}$]$_{0.59}$CoO$_{2}$, here used as a probe aimed at characterizing the coupling between 
itinerant and localized states. 
While a large negative magnetoresistance was already measured in sintered samples, \cite{Maignan03} 
we demonstrate here that it follows a scaling behavior with both magnetic field and temperature.
Originally introduced in the context of the colossal magnetoresistance of the manganites, \cite{Viret97,Wagner98} 
this scaling implies a spin dependent transport mechanism. 
Thus, we show that the analysis of the negative magnetoresistance allows to
shed new light on the electronic states leading to  high thermopower.

Similarly to Na$_{x}$CoO$_2$, the structure of the incommensurate cobalt oxide 
[Bi$_{1.7}$Ca$_{2}$O$_{4}$]$_{0.59}$CoO$_{2}$ (abbreviated thereafter BiCaCoO) contains single 
[CoO$_2$] layer of CdI$_2$ type stacked with rocksalt-type layers instead of a sodium deficient layer as displayed in Fig. \ref{figbicacoo}. 
One of the in-plane sublattice parameters being different from one layer to the other, \cite{Maignan03,Leligny99} the cobaltate BiCaCoO 
has a misfit structure as in most related compounds. \cite{Leligny99,Grebille07,Maignan03}
This system exhibits at room temperature a very large holelike thermopower S $\approx  $ 138  $\mu$V K$^{-1}$ 
and a rather low resistivity as in the so-called {\it bad} metals near a Mott metal-insulator transition. \cite{Maignan03,Georges96, Limelette03PRL}

\section{transport measurements}

\begin{figure}[htbp]
\centerline{\includegraphics[width=0.95\hsize]{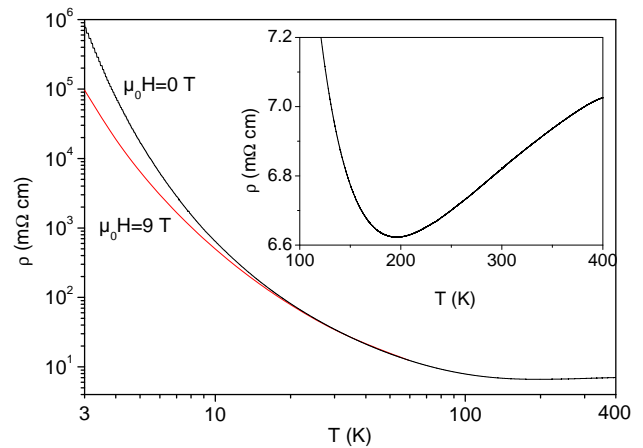}}
\caption{(color online). Temperature dependences of the single crystal in-plane resistivity $\rho$ at 0 T and 9 T, the magnetic field being parallel to the conducting plane. The inset displays a transport crossover from a high temperature metalliclike behavior to a low temperature insulatinglike one.
}
\label{fig0}
\end{figure}

In order to investigate the transport properties and more specifically the magnetoresistance in BiCaCoO, we have measured  the single crystal in-plane resistivity $\rho$(H,T) using a standard four terminal method. 
The studied single crystals were grown using a standard flux method, \cite{Leligny99} with typical sizes of the order of  4$\times$2$\times$0.02 mm$^3$.
As displayed in the inset of Fig.~\ref{fig0}, the temperature dependence of the resistivity exhibits a transport crossover around 200 K from a high temperature metalliclike behavior to a low temperature insulatinglike one. 
While in the former regime the values of the resistivity remain rather low, Fig.~\ref{fig0} displays a large enhancement of $\rho$ at low temperature indicating thus an efficient localization process. 
We also observe that the latter behavior is strongly reduced by the application of an in-plane magnetic field suggesting an unconventionnal insulatinglike regime.

\subsection{Magnetoresistance}

\begin{figure}[htbp]
\centerline{\includegraphics[width=0.95\hsize]{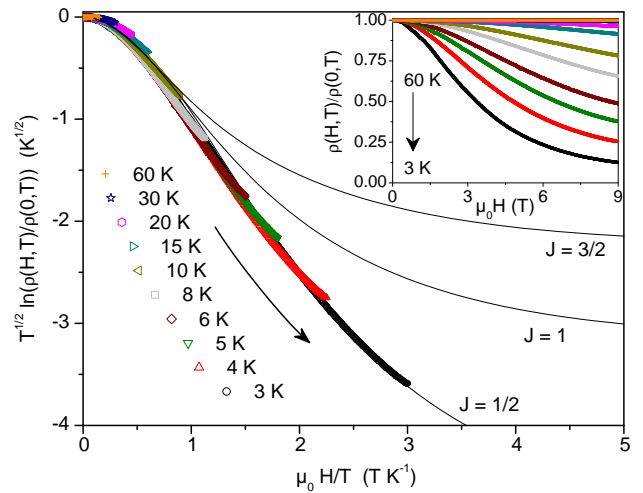}}
\caption{(color online). Scaling plot of the single crystal in-plane resistivity $\rho$(H,T) as a function of $\mu_0$H/T. 
The inset displays the magnetic field dependence of the normalized resistivity over the whole temperature range. 
The applied magnetic field is parallel to the conducting plane.
}
\label{fig1}
\end{figure}

In order to gain a better insight into this state, we have performed transport measurements as a function of an in-plane magnetic field H at constant 
temperature T. 
The data reported in the inset of Fig.~\ref{fig1} spans the T-range from 60 K down to 3 K with a large negative magnetoresistance reaching at the lowest temperature 85 $\% $ at 9 T. 
Also, Fig.~\ref{fig1} demonstrates that the whole set of magnetoresistance data can be scaled onto a variable range hopping form
(abbreviated thereafter VRH) which can be written as it follows.
\begin{equation}
\rho(H,T)=\rho_0(T) exp  \left( \frac{\tau_0(H/T)}{ T}\right)^{\alpha}
\label{eq1}
\end{equation}

While the two exponents $\alpha$=1/2 and 1/3 lead to satisfactory scalings, it is worth noting that the value $\alpha$=1/2 provides the best collapse of the experimental points onto a single curve. 
Originally introduced by Mott to describe the electronic conduction in disordered materials, the VRH theory essentially leads to the exponents $\alpha$=1/4 or 1/3 in three or two dimensions, and $\alpha$=1/2 in disordered systems with electron correlations. \cite{Mott79}

\begin{figure}[b!]
\centerline{\includegraphics[width=0.8\hsize]{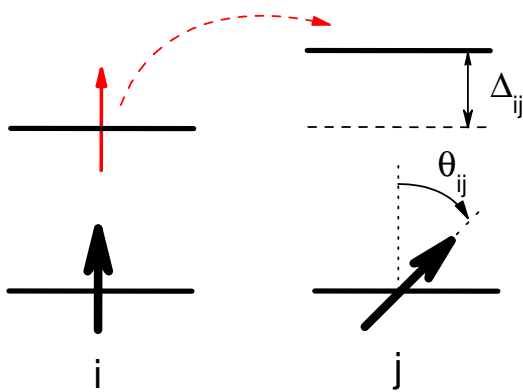}}
\caption{(color online). Schematic picture of the electronic energy levels involved in an elementary spin dependent hopping 
from the site i to the site j. 
The magnetic potential barrier $\Delta_{ij}$ results from the misorientation $\theta_{ij}$ between the quasiparticle spin $s_i$ 
(thin red arrow) and the localized spin $\sigma_j$ (bold black arrow) as explained in the text.
}
\label{fig2}
\end{figure}

More recently, the VRH mechanism has been extended to the case of magnetic disorder to explain the colossal negative 
magnetoresistance measured in manganites\cite{Viret97,Wagner98} and also applied in the Chromium based spinel compounds. \cite{muroi2001} 
The main ideas which underlie this conducting process consist in a variable range spin dependent hopping (VRSDH) 
due to the Hund's coupling -J$_H$\textbf{s}.\textbf{$\sigma$} between the quasiparticles spin \textbf{s} and the localized 
spin \textbf{$\sigma$}, with the Hund's coupling constant J$_H$. 
To basically illustrate this mechanism, let us consider the propagation of a quasiparticle 
with a spin s$_i$ from the site i to the site j as sketched in Fig.~\ref{fig2} with the involved spin polarized electronic energy levels.

By assuming a paramagnetic state, namely localized spins uncorrelated from one site to another, there is a probability to 
find an angle $\theta_{ij}$ between $\sigma_i$ and the neighbor localized spin $\sigma_j$. 
Because s$_i$ points along the localized spin $\sigma_i$ direction, the aforementioned misorientation leads to a magnetic 
potentiel barrier $\Delta_{ij}$=$\Delta$ (1-$\cos{\theta_{ij}}$)/2 with $\Delta$=2J$_H$ s$\sigma$.
At this very qualitative stage one can infer two important consequences which both agree with the experimental results reported in Fig.~\ref{fig1}.
First, the magnetic potential barriers are randomly distributed in such a paramagnetic material, namely there is a magnetic 
disorder which implies a VRH conduction. 
Besides, an applied magnetic field tends to align the localized spins and thus lowers both $\Delta_{ij}$ and the disorder, suggesting
a negative magnetoresistance.

In order to give a deeper insight into the spin dependent hopping that is more extensively developed in  Ref.~\onlinecite{Viret97}, 
an analogy can be performed with the conventional VRH process. 
In the latter case, the resistivity varies as $\rho\propto $exp(T$_0$/T)$^{\alpha}$ with T$_0 \propto$1/($\xi^3$ g(E$_F$)), $\xi$ 
being the localization length and g(E$_F$) the density of states at the Fermi level (DOS). \cite{Mott79}  
Without any magnetic field, the VRSDH resistivity should thus depend on a parameter $\tau_0\propto$1/($\xi^3$ n($\Delta_{ij}$))  
as in Eq.~(\ref{eq1}), with n($\Delta_{ij}$) the DOS of the magnetic potential barriers.  

Within this framework, the latter DOS can then be deduced from the probability of finding an angle $\theta_{ij}$ as  n($\Delta_{ij}$)$\approx$1/$\Delta$. 
Also, this constant DOS implies an average magnetic potentiel barrier $\langle \Delta_{ij}\rangle =\Delta/2$.  
The latter mean value is no longer valid when a magnetic field is applied because $\langle \cos{\theta_{ij}}\rangle \neq$0, and it 
must be written as $\langle \Delta_{ij}\rangle=\Delta (1-\langle \cos{\theta_{ij}}\rangle)/2$.  
In this case, the misorientation can be expressed from the local magnetizations $\vec{M}_{i,j}$ at the two sites i and j  
as $\langle\vec{M}_i \cdot \vec{M}_j\rangle$=M$_S^2$ $\langle$cos $\theta_{ij}\rangle$, with the saturation magnetization M$_S$. 
Since there are no short range correlations in a paramagnet, the averaged
magnetization is supposed to scale as the Brillouin function B$_J$(x) as
discussed by Wagner et al.\cite{Wagner98} It is given by:
\begin{equation}
B_J(x)=\frac{2J+1}{2J} \coth{\left( (2J+1) x\right)}- \frac{1}{2J} \coth (x)
\label{eqBj}
\end{equation}
Here $J$ is the angular momentum, $x=(g \mu_B H)/(2 k_BT)$, the Land\'e factor g=2, k$_B$ the 
Boltzmann constant and $\mu _B$ the Bohr magneton.  

As a consequence, one deduces that $\langle \cos{\theta_{ij}}\rangle$=B$_J$(x)$^2$ and thus the averaged magnetic potentiel
barrier is $\langle\Delta_{ij}\rangle$=$\Delta$ (1-B$_J$(x)$^2$)/2. 
Since it is lowered from its value without magnetic field by the factor (1-B$_J$(x)$^2$), one may infer that the new DOS 
is enhanced as n($\Delta_{ij}$)$\approx$1/($\Delta$(1-B$_J$(x)$^2$) by assuming that it keeps constant over the whole range of potential. 
In this approximation, the temperature $\tau_0$(x) can be written following Eq.~(\ref{eqtau}) with the unit cell volume $v$.

\begin{equation}
\tau_0(x)= \left(1-B_J(x)^2 \right) T_0^M    \makebox{\ and \ } T_0^M \approx
\frac{v}{\xi^3} \frac{\Delta}{ k_B}
\label{eqtau}
\end{equation}

The combination of both Eq.~(\ref{eq1}) and Eq.~(\ref{eqtau}) leads to a complete analytic expression of the magnetoresistance 
with the only two parameters J and T$_0^M$ that yields to the curves in Fig.~\ref{fig1} compared to the experimental data. 
The three tested sets of parameters with J=1/2 (T$_0^M$=24 K), J=1 (T$_0^M$=12 K) and J=3/2 (T$_0^M$=6 K) 
demonstrate unambiguously the agreement with J=1/2 by also confirming the relevance of the analysis. 
One can here emphasize that the found rather low value of T$_0^M$ is consistent with a VRH transport since it suggests 
as required a localization length exceeding the interatomic distances. 
Beyond these strong checks, this result implies that the localized spin involved here in the VRSDH mechanism in BiCaCoO is $\sigma$=1/2 
as previously inferred from the analysis of the magnetic field dependence of the thermopower\cite{LimelettePRL06} and 
in agreement with the previously reported susceptibility measurements. \cite{Maignan03}
As a consequence, it seems that both extended and localized states not only coexist but also give rise to unusual electronic 
properties as large negative magnetoresistance and high thermopower.

\begin{figure}[htbp]
\centerline{\includegraphics[width=0.95\hsize]{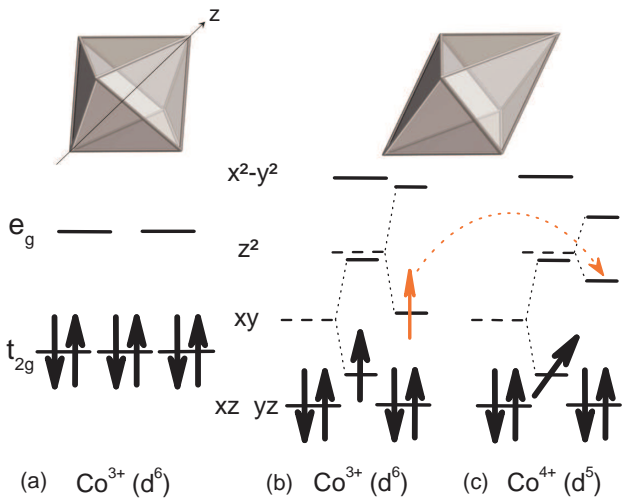}}
\caption{(color online). (a): Crystal field effect in the Co$^{3+}$ (d$^6$).
(b): The structural distorsion along the z axis illustrated by the two upper octahedrons reduces the gap between 
the d$_{z^2}$ and the d$_{xy}$ levels. 
A strong Hund's coupling is besides represented with the splitting of the two spin states in the d$_{z^2}$ and d$_{xy}$ levels respectively.
(b) and (c): Spin dependent hopping mechanism between the Co$^{3+}$ and the Co$^{4+}$. 
The thin red arrow represents the quasiparticle spin and the bold black arrows the localized spin.}
\label{fig3}
\end{figure}

Let us now discuss the relevance of the electronic levels as sketched in Fig.~\ref{fig2} in BiCaCoO. 
By analogy with the crystallographic structure of  Na$_x$CoO$_2$,  \cite{singh2000} the [CoO$_2$] planes in the misfit cobaltates 
are assumed to be conducting and are stacked with the insulating rocksalt-type layers. 
These [CoO$_2$] layers consist in a 2D triangular lattice Co sheets octahedrally coordinated with O above and below 
the Co plane. 
As depicted in Fig.~\ref{fig3}, the Co d levels are first crystal field split in the octahedral O environment into a lower lying t$_{2g}$ and 
a upper lying e$_g$ manifold. 
In this context, the six electrons of the Co$^{3+}$ (d$^6$) occupy the t$_{2g}$ levels (see Fig.~\ref{fig3}(a)). 
Second, the  difference in energy between the d$_{z^2}$ and d$_{xy}$ levels can be reduced assuming a structural distortion of the octahedron following
from a stretching of the the z axis. 
Therefore, for sufficiently strong Hund's coupling, the gap between the d$_{z^2}$ and the d$_{xy}$ levels can become smaller than the Hund's rule splitting,  
\textit{i.~e.} the spin up d$_{z^2}$ energy can get smaller than the spin down d$_{xy}$ energy. As a result the scattering of itinerant e$_g$ electrons
hopping from site to site by the localized t$_{2g}$ electrons, as illustrated in Fig.~\ref{fig3}(b) and Fig.~\ref{fig3}(c), leads to the same electronic
background as in Fig.~\ref{fig2}, and explains the above-mentioned negative magnetoresistance. 

Even if this situation can be achieved from others scenarios, possibly involving others orbitals instead of the d$_{z^2}$ and the d$_{xy}$ ones, we emphasize that the aforementioned ingredients must remain to explain the magnetoresistance,  namely a distorsion and a Hund's splitting which reduce the t$_{2g}$-e$_{g}$ gap and spin polarize the electronic levels.
Also, it is worth noting that a similar description with a strong coupling
between the spin and the orbital degrees of freedom has been quite recently
proposed in order to explain the magnetic properties in the parent compound
Na$_x$CoO$_2$.  \cite{daghofer2006}

\subsection{Thermopower}

\begin{figure}[htbp]
\centerline{\includegraphics[width=0.95\hsize]{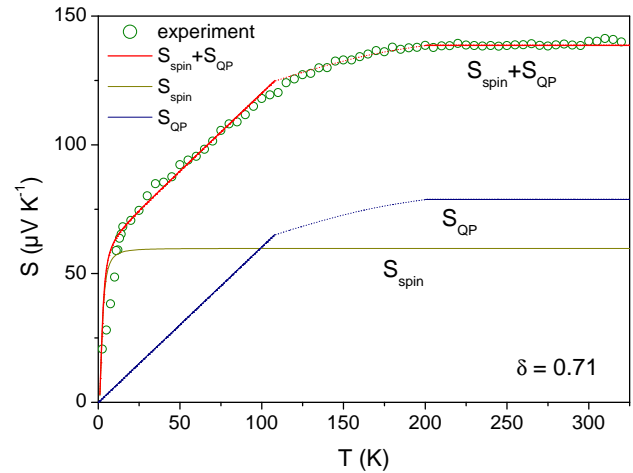}}
\caption{(color online). Comparison between the experimental temperature dependence of the thermopower $S$, \cite{Maignan03} 
and the theoretical predictions including both localized spins 
$S_{\rm spin}$, and quasiparticles $S_{\rm QP}$ contributions. 
It is worth noting that the latter is plotted from the two limiting cases,
  namely with $S_{\rm QP}\approx \pi^2$/6 k$_B$/e T/T$^*_F$ 
if T$<<$T$^*_F$ and $S_{\rm QP}$= k$_B$/e ln($\delta$/(1-$\delta$)) if
  T$>$T$^*_F$. 
}
\label{fig4}
\end{figure}

We now would like to address more specifically the connection between the electronic levels represented in Fig.~\ref{fig3}(b) 
and \ref{fig3}(c) and the thermopower S measured in BiCaCoO and reported in Fig.~\ref{fig4}. \cite{LimelettePRL06}

According to Fig.~\ref{fig3}(b) and \ref{fig3}(c), the t$_{2g}$ electrons are localized ($\sigma$=1/2), while the lower e$_g$ electron (s=1/2)
are itinerant, preventing them from forming local triplet states on long time scales. Since the localized spins $\sigma$ remain paramagnetic, they
asymptotically give rise to a thermopower contribution $S_{\rm spin}$=k$_B$/e ln(2) due to their spin entropy as displayed in Fig.~\ref{fig4}, with $e$ the
elementary electric charge. 
In order to provide a constant contribution over a wide temperature range, it is here assumed that their energy level is located in an energy width of a few
 tens of kelvin below the Fermi level and that it does not hybridize with the e$_g$ band. 
We stress that the latter interpretation is consistent with the fact that these spins are also responsible for a negative magnetothermopower. \cite{LimelettePRL06}

On the other hand, the e$_g$ quasiparticles are expected to contribute to the thermopower as $S_{\rm QP}\propto$ T/T$^*_F$ at  temperatures lower than the effective Fermi temperature $T^*_F\approx$T$_F$/m$^*$. 
Here T$_F$ is the bare Fermi temperature and m$^*$ is the effective mass which takes into account the electronic correlations. \cite{Palsson98,Merino00} 
Note that, according to P\'alsson and Kotliar, \cite{Palsson98} the thermopower does not depend on the scattering rate in the low temperature regime.
Also, a rough estimate of the effective Fermi temperature inferred from the slope of S leads to a very low value ($230~K < T^*_F < 260~K$) 
which implies an enhanced m$^*$ in agreement with the observed constant thermopower for T$>$T$^*_F$. 
In fact, at  temperatures higher than T$^*_F$ the e$_g$ electronic excitations become incoherent and the related thermopower recovers a 
purely entropic form following Eq.~(\ref{eqSQP}). 
In this regime, the thermopower is temperature independent and reads
\begin{equation}
S_{\rm QP}( T>T^*_F)= \frac{k_B}{e} ln \left( \frac{\delta}{N(1-\delta)} \right)
\label{eqSQP}
\end{equation}
with the degeneracy N involving both spin and orbital degrees of freedom \cite{Palsson98} and the electronic filling factor $\delta$.
Moreover, because of their spin polarized level the e$_g$ electrons have no more spin degeneracy, implying thus N=1 in Eq.~(\ref{eqSQP}). 
By adding the two aforementioned entropic contributions in Fig.~\ref{fig4}, namely those originating from both the paramagnetic spins 
(the t$_{2g}$ electrons) and the quasiparticles (the e$_g$ electrons), one can therefore deduce the filling factor $\delta \approx $ 0.71 
corresponding to 0.29 hole.
Let us finally discuss the electronic properties of the parent compound BiBaCoO. \cite{Hervieu2003}
This cobaltate exhibits neither negative magnetoresistance nor negative magnetothermopower. 
The low temperature linear in T dependence of S extrapolates to zero in contrast to Fig.~\ref{fig4}, suggesting that only extended states 
contribute to the thermopower in this compound. 
Also, the room temperature value of the thermopower is of the order of 90 $\mu$V K$^{-1}$, namely 50 $\mu$V K$^{-1}$ less than in BiCaCoO.
As a strong check of the whole reported analysis, this comparison clearly demonstrates, by invalidating the presence 
of localized paramagnetic spins in BiBaCoO, that these dual electronic states are a source of an enhanced thermopower.
Further experimental investigations are now needed in order to understand why these coexisting states are not observed in BiBaCoO 
in contrast to BiCaCoO.

\section{conclusion}

To conclude, we have investigated the low temperature magnetic field dependence of the resistivity in the thermoelectric misfit 
cobalt oxide BiCaCoO from 60 K down to 3 K. 
The scaling of the negative magnetoresistance demonstrates a spin dependent transport mechanism due to a strong Hund's coupling. 
The inferred microscopic description implies dual electronic states which explain the coexistence between localized and itinerant electrons 
both contributing to the thermopower.
By shedding a new light on the electronic states leading to a high thermopower, this result likely provides a new potential way to optimize the thermoelectric properties.

\begin{acknowledgments}
We are grateful to F. Ladieu and D. Grebille for illuminating discussions.
\end{acknowledgments}

%\bibliography{cobaltate}

\end{document}